\documentclass[aps,preprint]{revtex4-1}
\usepackage{graphicx}
\usepackage{amsmath}
\usepackage{makeidx}
\usepackage{amsfonts}
\usepackage{amssymb}

\begin{document}

\title{Sign problem in the Bethe approximation}

\author{A. Ramezanpour}
\email{abolfazl.ramezanpour@polito.it}
\affiliation{Physics Department and Center for Computational Sciences, Politecnico di Torino, C.so Duca degli Abruzzi 24, I-10129 Torino, Italy}

\author{R. Zecchina}
\email{riccardo.zecchina@polito.it}
\affiliation{Physics Department and Center for Computational Sciences, Politecnico di Torino, Corso Duca degli Abruzzi 24, 10129 Torino, Italy}
\affiliation{Human Genetics Foundation, Torino, via Nizza 52, 10126 Torino, Italy}
\affiliation{Collegio Carlo Alberto, Via Real Collegio 30, 10024 Moncalieri, Italy}

\date{\today}

\begin{abstract}
We propose a message-passing algorithm to compute the Hamiltonian expectation with respect to an appropriate class of trial wave functions for an interacting system of fermions. To this end, we connect the quantum expectations to average quantities in a classical system with both local and global interactions, which are related to the variational parameters and use the Bethe approximation to estimate the average energy within the replica-symmetric approximation. The global interactions, which are needed to obtain a good estimation of the average fermion sign, make the average energy a nonlocal function of the variational parameters. We use some heuristic minimization algorithms to find approximate ground states of the Hubbard model on random regular graphs and observe significant qualitative improvements with respect to the mean-field approximation.
\end{abstract}


\maketitle

\section{Introduction}\label{S0}
Finding the ground-state of a quantum system can be recast as an optimization problem by minimizing the Hamiltonian
expectation over the space of trial wave functions. In practice, it is important for the efficiency of the variational method,
to have a succinct representation of the trial wave functions that accurately describe the ground-state of the quantum system \cite{MC-rmp-2001,VM-prl-2005,H-prb-2006}. Nevertheless, finding the optimal variational parameters could be a hard computational task even in one dimension \cite{AGIK-mphys-2009} since the objective function is an average quantity computed over the exponentially large Hilbert space of the physical system, let alone the complex landscape of the energy function induced by different sources of frustrations. And the problem is more serious for fermions due to the global nature of the fermion sign \cite{TW-prl-2005}. Still, the main strategy to deal with the above variational problem, is to use Monte Carlo (MC) method both in computing the Hamiltonian expectation and in optimizing over the variational parameters \cite{S-prb-2007}. 

In this paper we further develop the variational quantum cavity method introduced in Ref. \cite{R-prb-2012} to study the ground-state properties of an interacting fermion system. 
More precisely, for a given instance of the variational parameters, we connect the quantum expectations to average quantities in a classical system of interacting particles or spins, where the interactions are related to the variational parameters. Then, instead of MC sampling, we use the Bethe approximation \cite{B-prs-1935}, or cavity method in the replica-symmetric approximation \cite{MPV-book-1987,MP-epjb-2001} to estimate the classical expectations. Within the Bethe approximation, the probability marginals are obtained by an efficient and local message-passing (MP) algorithm \cite{KFL-inform-2001,BMZ-rsa-2005}; the estimated marginals are good as long as the interaction graph is locally tree-like, the classical system is in a replica-symmetric phase and it is effectively mean-field \cite{MM-book-2009}. Some applications of the cavity method in quantum systems can be found in Refs. \cite{H-prb-2007,LP-aphys-2008,LSS-prb-2008,KRSZ-prb-2008,STZ-prb-2008,IM-prl-2010}. One may find some connections among these papers, the (statistical) dynamical mean-field theory \cite{DK-prl-1997}, and density-matrix renormalization group \cite{W-prl-1992}.        

The trial wave functions can be characterized by the type of interactions included in the associated classical system. We usually start from 
the mean-field (MF) approximation considering only the one-body interactions, and improve on that by adding higher-order interactions to capture the relevant correlations. For bosons, a good estimation of the quantum expectations can be obtained by considering local interactions involving only a few number of particles \cite{VMB-2008}. As a result, the Hamiltonian expectation is a local function of the variational parameters and we can again utilize the Bethe approximation to estimate the optimal parameters by a higher level MP algorithm \cite{R-prb-2012,ABRZ-prl-2011}.  In the case of fermions, however, we have to work with global interactions involving an extensive number of particles to deal with the global nature of the fermion sign. Consequently, the average energy becomes a nonlocal function of the variational parameters and we can not exploit the local MP algorithms to optimize over the parameters anymore. 

In this paper we take the Hubbard model and propose a class of trial wave functions with both local and global interactions, where the Hamiltonian expectation can be computed by an MP algorithm. Using some heuristic minimization algorithms we find approximate ground-states of the Hubbard model in random regular graphs of degree $C=3$. The results are considerably better than the MF predictions, and are close to the exact solutions in small systems. For comparison we also present some results in one- and two-dimensional lattices. 
 
In the next section we give some definitions and use the mean-field approximation to illustrate the main points that are relevant for the following discussions. In Sec. \ref{S2} we introduce the global ansatz of the wave functions and the machinery we need to deal with the global interactions within the Bethe approximation. The numerical data are presented in Sec. \ref{S3} and finally we summarize the results in Sec. \ref{S4}.

\section{Hubbard model in the mean-field approximation}\label{S1}
Consider the Hubbard model with Hamiltonian $H=H_0+H_1$ where 
\begin{align}
H_0 &=\sum_{i} U_{i}c_{i \uparrow}^{\dagger}c_{i \uparrow}c_{i \downarrow}^{\dagger}c_{i \downarrow}-\sum_{i} \nu_i (c_{i \uparrow}^{\dagger}c_{i \uparrow}+c_{i \downarrow}^{\dagger}c_{i \downarrow}),\\ \nonumber
H_1 &=-\sum_{(ij) \in \mathcal{E}_q,\sigma} t_{ij}(c_{j \sigma}^{\dagger}c_{i \sigma}+c_{i \sigma}^{\dagger}c_{j \sigma}),
\end{align} 
with index $i=1,\dots,N$ that labels the sites in the quantum interaction graph $\mathcal{E}_q$. 
The $c_{i\sigma}^{\dagger}$ and $c_{i\sigma}$ are creation and annihilation operators for a fermion of spin $\sigma=\uparrow, \downarrow$ at site $i$.
We will work in the occupation number representation  $|\underline{n} \rangle=
(c_{1\uparrow}^{\dagger})^{n_{1\uparrow}}\dots (c_{N\uparrow}^{\dagger})^{n_{N\uparrow}}
(c_{1\downarrow}^{\dagger})^{n_{1\downarrow}}\dots (c_{N\downarrow}^{\dagger})^{n_{N\downarrow}} |0\rangle $
and will take the following order of the sites and spins: $(1 \uparrow,  \dots, N \uparrow)(1 \downarrow,  \dots, N \downarrow)$. We assume that
there is a path in $\mathcal{E}_q$ connecting $1\to 2 \to \dots \to N$ representing the ordering backbone.  
Given a trial wave function $|\Psi(\underline{P})\rangle=\sum_{\underline{n}} \psi(\underline{n};\underline{•}e{P}) |\underline{n} \rangle$ depending on a set of variational parameters $\underline{P}$, we write
the Hamiltonian expectation as $\langle \Psi(\underline{P}) | H |\Psi(\underline{P})\rangle= \sum_{\underline{n}} |\psi(\underline{n};\underline{P})|^2[ E_0(\underline{n})+E_1(\underline{n})]$ with
\begin{align}
E_0(\underline{n})&\equiv \langle \underline{n} | H_0 |\underline{n} \rangle \equiv \sum_i e_i(n_{i\uparrow},n_{i\downarrow}), \\ 
E_1(\underline{n})&\equiv \mathrm{Re}\left\{ \sum_{\underline{n}'} \frac{\psi^*(\underline{n}';\underline{P})}{\psi^*(\underline{n};\underline{P})} \langle \underline{n}' | H_1 |\underline{n} \rangle \right\} \equiv \sum_{(ij) \in \mathcal{E}_q,\sigma} (-1)^{\sum_{i<k<j} n_{k\sigma}} e_{ij\sigma}(\underline{n}).
\end{align}
Depending on the trial wave function we obtain different expressions for $e_{ij\sigma}(\underline{n})$ but we always have   
$e_{i}(n_{i\uparrow},n_{i\downarrow}) = U_{i} n_{i\uparrow}n_{i\downarrow}-\nu_i (n_{i\uparrow}+ n_{i\downarrow})$.

The goal is to consider $\mu(\underline{n};\underline{P})=|\psi(\underline{n};\underline{P})|^2$ as a probability measure 
in a classical system and to compute the above average quantities within the Bethe approximation.
The classical measure is, in general, represented by  
$\mu(\underline{n};\underline{P}) \propto 
\prod_{a} \phi_a(\underline{n}_{\partial a})$
with the set of classical interactions  
$\mathcal{E}_c\equiv \{\phi_a(\underline{n}_{\partial a})|a=1,\dots,A\}$. 
Here, $\partial a$ is the subset of variables that appear in interaction $a$. 

In a MF approximation, we take a factorized trial wave function including the Gutzwiller interactions \cite{G-prl-1963},
\begin{align}
\psi(\underline{n};\underline{P}) \propto \prod_i \exp\left( K_{i} n_{i\uparrow}n_{i\downarrow}+\sum_{\sigma} B_{i\sigma} n_{i\sigma} \right),
\end{align} 
with complex parameters $K_i, B_{i\uparrow}$, and $B_{i\downarrow}$. 
This results in the following classical measure 
$\mu(\underline{n};\underline{P}) \propto \prod_{i} \exp\left( 2K_{i}^R n_{i\uparrow}n_{i\downarrow}+\sum_{\sigma} 2B_{i\sigma}^R n_{i\sigma}\right)\equiv \prod_i \mu_i(n_{i\uparrow},n_{i\downarrow})$.
By superscript $R$, we mean the real part of the parameters.
Given the above measure, we find   
\begin{align}
e_{ij\sigma}(\underline{n}) &= -t_{ij} 
\mathrm{Re}\left\{  \delta_{n_{j\sigma}n_{i\sigma},01}
e^{-\Delta_{n\sigma} \psi }
+\delta_{n_{i\sigma}n_{j\sigma},01}
e^{+\Delta_{n\sigma} \psi }  \right\},
\end{align}
where we defined
$\Delta_{n\sigma} \psi=(K_i^* n_{i\bar{\sigma}}-K_j^* n_{j\bar{\sigma}})+(B_{i\sigma}^*-B_{j\sigma}^*)$ with $\bar{\uparrow}=\downarrow$ and vice versa.
Here we can easily compute the average local energies, e.g.,
\begin{align}
\langle (-1)^{\sum_{i<k<j}n_{k\sigma}}e_{ij\sigma}(\underline{n}) \rangle_{\mu} &= -t_{ij}  \frac{2}{Z_iZ_j}(R_{i\sigma}R_{j\sigma}+S_{i\sigma}S_{j\sigma}) \prod_{i<k<j}\left[\mu_k(n_{k\sigma}=0)-\mu_k(n_{k\sigma}=1)\right],
\end{align}
with $Z_i = e^{2K_i^R+2B_{i\uparrow}^R+2B_{i\downarrow}^R}+e^{2B_{i\uparrow}^R}+e^{2B_{i\downarrow}^R}+1$ and
\begin{align}
R_{i\sigma} &= e^{K_i^R+B_{i\sigma}^R+2B_{i\overline{\sigma}}^R}\cos(K_{i}^I+B_{i\sigma}^I)+e^{B_{i\sigma}^R}\cos(B_{i\sigma}^I), \\
S_{i\sigma} &= e^{K_i^R+B_{i\sigma}^R+2B_{i\overline{\sigma}}^R}\sin(K_{i}^I+B_{i\sigma}^I)+e^{B_{i\sigma}^R}\sin(B_{i\sigma}^I).
\end{align}
There are a few points to mention here: First, the only difference with a bosonic system is the sign term 
$(-1)^{\sum_{i<k<j}n_{k\sigma}}$. It is clear that, 
in the absence of this sign and for $t_{ij}\ge 0$, we can minimize the average energies by setting the imaginary part of the parameters to zero.
When the sign term is present or the $t_{ij}$ take different signs, one can show that, starting from real parameters, one always remains 
in the real subspace of the parameters following a gradient descent algorithm. This is true not only for the MF wave function, but also for the class of wave functions that we consider in this paper.   
Second, in the
MF approximation, the average of the sign term is exponentially small in the number of sites $k$ between $i$ and $j$. This suggests that smaller average energies are
obtained by maximizing the overlap between the ordering chosen in the trial wave function and the quantum interaction graph $\mathcal{E}_q$. 
Moreover, the density profile would also depend on the ordering unless the parameters are constrained
to respect the system's translational symmetries.   
This artifact of the MF approximation has to be cured by adding interactions to the classical interaction graph $\mathcal{E}_c$ 
to correlate distant variables along the ordering backbone. 
And finally, due to the sign term the average energy is not a local function of the parameters. This sets some restrictions on the optimization
algorithms that we can use in minimizing the Hamiltonian expectation.

\section{Beyond the mean-field approximation: Local and global interactions}\label{S2}
The simplest interactions to improve the MF approximation are local two-body or Jastrow interactions $J_{ij\sigma}n_{i\sigma}n_{j\sigma}$ \cite{J-pr-1955}. It is not difficult to guess that these interactions are not enough to capture the sign correlations. The interaction set could, of course, be enlarged by adding other many-body interactions also including different types of spins. Instead, here, we take another approach by introducing global variables $\xi_{i\sigma}\equiv (-1)^{\sum_{k=1}^i n_{k\sigma}}$. Then the sign term $(-1)^{\sum_{i<k<j}n_{k\sigma}}$ can be written as $\xi_{i\sigma}\xi_{j\sigma}(-1)^{n_{j\sigma}}$, which is a local function of the global variables \cite{steiner-prl-2008,RRZ-epjb-2011}. Accordingly, we 
can have global one-body interactions $\Theta_{i\sigma} \xi_{i\sigma}$ and global two-body interactions $\Gamma_{ij\sigma}\xi_{i\sigma}\xi_{j\sigma}$ in the classical interaction graph. We call this set of trial wave functions the global ansatz. In general one could have interactions of type $J_{i_1\dots i_mj_1\dots j_{m'}} n_{i_1\sigma_{i_1}}\dots n_{i_m\sigma_{i_m}} \xi_{j_1\sigma_{j_1}}\dots \xi_{j_{m'}\sigma_{j_{m'}}}$. 

Notice that the above interactions do not necessarily respect the symmetries of the system.  However, by minimizing the Hamiltonian expectation over the variational parameters, we get closer to the ground state of the system and, therefore, minimizing the effect of these asymmetries.

In the following, we consider the global one- and two-body interactions, i.e.:
\begin{align}
\psi(\underline{n};\underline{P}) &\propto \exp\left(\sum_{i} K_{i} n_{i\uparrow}n_{i\downarrow}+\sum_{i\sigma}B_{i\sigma} n_{i\sigma}+\sum_{i\sigma}\Theta_{i\sigma}\xi_{i\sigma}+\sum_{(ij)\in \mathcal{E}_c, \sigma}\Gamma_{ij\sigma}\xi_{i\sigma}\xi_{j\sigma} \right),
\end{align}
where, for simplicity, we are going to assume $\mathcal{E}_c=\mathcal{E}_q$.
As a result, we obtain 
\begin{align}
e_{ij\sigma}(\underline{n},\underline{\xi}) =  -t_{ij} \mathrm{Re}\left\{ \delta_{n_{j\sigma}n_{i\sigma},01}
e^{ -\Delta_{n\sigma} \psi - \Delta_{\xi \sigma} \psi } + \delta_{n_{i\sigma}n_{j\sigma},01}e^{ +\Delta_{n\sigma} \psi - \Delta_{\xi \sigma} \psi } \right\},
\end{align}
with $\Delta_{n\sigma} \psi$ as given before and 
\begin{align}
\Delta_{\xi \sigma} \psi =\sum_{i \le k <j} 2\Theta_{k \sigma}^*\xi_{k\sigma}+ \sum_{(kl): k< i, i\le l < j} 2\Gamma_{kl \sigma}^*\xi_{k\sigma}\xi_{l\sigma}
+\sum_{(kl): i\le k< j,  l\ge j} 2\Gamma_{kl \sigma}^*\xi_{k\sigma}\xi_{l\sigma}.
\end{align}
The average energy is computed with respect to the following classical measure: $\mu(\underline{X}) \propto \prod_i \phi_i(X_i)\prod_{(ij)\in \mathcal{E}_c} \phi_{ij}(X_i,X_j)$ where, for brevity, we defined $X_i\equiv \{n_{i\uparrow},\xi_{i\uparrow},n_{i\downarrow},\xi_{i\downarrow}\}$, and
\begin{align}
\phi_i(X_i) &\equiv I_{\xi_i} e^{ 2K_{i}^R n_{i\uparrow}n_{i\downarrow}+\sum_{\sigma} 2B_{i\sigma}^R n_{i\sigma}+\sum_{\sigma} 2\Theta_{i\sigma}^R\xi_{i\sigma}},\\
\phi_{ij}(X_i,X_j) &\equiv I_{\xi_i,\xi_j} e^{\sum_{\sigma}2\Gamma_{ij\sigma}^R\xi_{i\sigma}\xi_{j\sigma}}.
\end{align}
The indicator functions $I_{\xi_i,\xi_j}$ check the constraints $\xi_{i\sigma}=(-1)^{n_{i\sigma}} \xi_{(i-1)\sigma}$ on the global variables if $j=i\pm 1$, and $I_{\xi_i}$ fixes the boundary condition $\xi_{1\sigma}=(-1)^{n_{1\sigma}}$ when $i=1$. To estimate the average energy, we resort to the Bethe approximation, writing the local marginals in terms of the cavity ones satisfying the following set of equations \cite{MM-book-2009}:  
\begin{align}
\mu_{i \to j}(X_i) \propto  \phi_i(X_i)\prod_{k \in \partial i \setminus j}\left( \sum_{X_k} \phi_{ik}(X_i,X_k) \mu_{k \to i}(X_k) \right), 
\end{align}
where $\partial i$ denotes the neighborhood set of site  $i$ in $\mathcal{E}_c$.  These are the 
belief propagation (BP) equations \cite{KFL-inform-2001} that are solved by iteration starting from random initial cavity marginals. In the replica-symmetric approximation we assume there is a fixed point to the BP equations describing the single Gibbs state of the system. 
The average $\langle e_i(n_{i\uparrow},n_{i\downarrow}) \rangle_{\mu}$ is simply computed after the local marginal $\mu_i(X_i)$, which is 
computed like $\mu_{i \to j}(X_i)$ but taking all the neighbors into account.      
In the other part of the average energy, we need to compute expectations, such as $\langle \xi_{i\sigma}\xi_{j\sigma}(-1)^{n_{j\sigma}}\delta_{n_{j\sigma}n_{i\sigma},01} \exp\left( -\Delta_{n\sigma} \psi - \Delta_{\xi \sigma} \psi \right) \rangle_{\mu}$. To get around the problem of computing the average of a global quantity we rewrite it as 
$\exp(F-\tilde{F})\langle \xi_{i\sigma}\xi_{j\sigma}(-1)^{n_{j\sigma}}\delta_{n_{j\sigma}n_{i\sigma},01} 
e^{-\Delta_{n\sigma} \psi}\rangle_{\tilde{\mu}}$, introducing the complex measure $\tilde{\mu}(\underline{X}) \propto \exp\left(- \Delta_{\xi \sigma} \psi \right) \mu(\underline{X})$ and the corresponding free energy $\tilde{F}$. The free energy difference in the Bethe approximation is given by $F-\tilde{F}= \sum_i(\Delta F_i-\Delta \tilde{F}_i)-\sum_{(ij)\in \mathcal{E}_c}(\Delta F_{ij}-\Delta \tilde{F}_{ij})$, where $\Delta F_i$ and $\Delta F_{ij}$ are the free-energy changes by adding site $i$ and the interaction between sites $i$ and $j$, respectively, that is,
\begin{align}
e^{-\Delta F_i} &= \sum_{X_i} \phi_i(X_i) \prod_{k \in \partial i}\left( \sum_{X_k} \phi_{ik}(X_i,X_k) \mu_{k \to i}(X_k)\right), \\
e^{-\Delta F_{ij}} &= \sum_{X_i,X_j}  \phi_{ij}(X_i,X_j) \mu_{i \to j}(X_i)
\mu_{j \to i}(X_j),
\end{align}
and similarly for the complex measure \cite{MM-book-2009}. In this way, we can compute the Hamiltonian expectation for the above class of trial wave functions
with a local message-passing algorithm in a time complexity of order $N^2$ for sparse classical and quantum interaction graphs. Note that a small error in estimating the classical free energies could result in a large error in estimating the average energy due to the exponential factor $\exp(F-\tilde{F})$.

\section{Numerical simulations}\label{S3}
Having the Hamiltonian expectation for an arbitrary instance of the variational parameters, we need an optimization algorithm to
find the optimal parameters. This is a computationally hard problem, and we have to resort to some heuristic algorithms to
find an approximate ground state for the system. Let us start from a local minimization algorithm where,
in each step, we fix all the parameters except in a small region of the system and minimize the associated energy contribution.
For instance, in case $\Gamma_{ij\sigma}=0$, we take the subset $\{K_i, B_{i\uparrow}, B_{i\downarrow},\Theta_{i\uparrow}, \Theta_{i\downarrow}\}$ and minimize the following energy:
\begin{equation}
E_{r}\equiv \langle e_i \rangle_{\mu}+\sum_{j \in \partial i}\sum_{\sigma}
\langle \xi_{i\sigma}\xi_{j\sigma} (-1)^{n_{\max(i,j)\sigma}}e_{ij}\rangle_{\mu}
 +\sum_{(kl): k<i<l}\sum_{\sigma}\langle \xi_{k\sigma}\xi_{l\sigma} (-1)^{n_{\max(k,l)\sigma}}e_{kl}\rangle_{\mu}.
\end{equation}
The energy function is chosen as the sum of the average energies that explicitly depend on the subset of the parameters.
The index $i$ is selected randomly, and the corresponding parameters are updated. The process ends when no local update can decrease the average energy. 

In another algorithm, we use a population of the parameters $\{\underline{P}^a|a=1,\dots,N_p\}$ and update the population in order to find smaller average energies. More precisely, in each step, we select two sets of parameters $(\underline{P}^a,\underline{P}^b)$ and find the set $\underline{P}^{ab}$ minimizing the average energy along the line $\lambda \underline{P}^a+(1-\lambda)\underline{P}^b$ for $\lambda\in [0,1]$. Then, we replace the maximal member of the population with $\underline{P}^{ab}$ and change the position of points $a$ and $b$ to somewhere between $\underline{P}^{ab}$ and the minimal member of the population $\underline{P}^{min}$. That is,  $\underline{P}^a=(\underline{P}^{ab}+\underline{P}^{min})/2+\underline{R}^a$ and $\underline{P}^b=(\underline{P}^{ab}+\underline{P}^{min})/2+\underline{R}^b$ for some random vectors $(\underline{R}^a,\underline{R}^b)$.

And finally, in a gradient descent algorithm, the parameters are updated as $\delta P=-\eta \frac{\partial \langle H \rangle_{\mu}}{\partial P}$ for some small and positive $\eta$'s. This means that we need to have the cavity susceptibilities $\chi_{i \to j}^{P}(X_i)\equiv \frac{\partial \ln \mu_{i \to j}(X_i)}{\partial P}$, which can be obtained by taking the derivative of the BP equations. For instance, we have
\begin{align}
\chi_{i \to j}^{K_l}(X_i)=n_{l\uparrow}n_{l\downarrow}\delta_{l,i}+\sum_{k \in \partial i \setminus j} \langle \chi_{k \to i}^{K_l}(X_i) \rangle_{k\to i}-C_{i \to j}^{K_l}, 
\end{align}
if $I_{\xi_i}=1$, otherwise, $\chi_{i \to j}^{K_l}(X_i)=0$. 
Here we defined
\begin{align}
\langle \chi_{k \to i}^{K_l}(X_i) \rangle_{k\to i} \equiv \frac{ \sum_{X_k} \phi_{ik}(X_i,X_k) \mu_{k \to i}(X_k)\chi_{k \to i}^{K_l}(X_k)}
{\sum_{X_k} \phi_{ik}(X_i,X_k) \mu_{k \to i}(X_k) }, 
\end{align}
and $C_{i \to j}^{K_l}$ is obtained by normalization $\sum_{X_i} \mu_{i \to j}(X_i) \chi_{i \to j}^{K_l}(X_i)=0$.
These are called the susceptibility propagation equations \cite{MM-sp-2009}. Similarly, we can write the equations for the cavity susceptibilities for the complex measure defined in the previous section.

We can use a combination of the above algorithms to approach the optimal parameters, for example, the local minimization algorithm followed by the gradient descent algorithm. In the following, we always start from zero initial parameters or a population of parameters distributed randomly around zero in the case of population dynamics. The algorithms are repeated a number of times to get the best outcome for different realizations of the update process. 

\begin{figure}
\includegraphics[width=10cm]{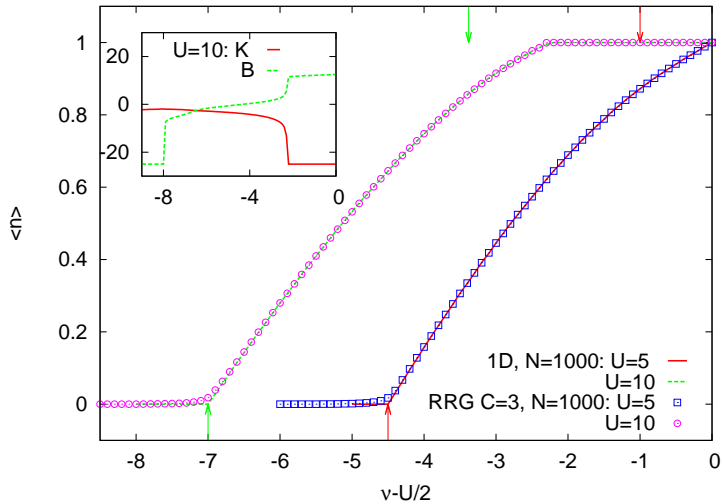}
\caption{The average density in the mean-field approximation for the homogeneous Hubbard model in a random regular graph (RRG) of degree $C=3$ and in one dimension (1D). The data are obtained by the local minimization algorithm followed by the gradient descent algorithm and repeated for $10$ different realizations of the update process. The arrows show the expected phase transition points for the one-dimensional system in the thermodynamic limit. The inset shows the optimal parameters when the wave function is given by $K_i=K, B_{i\uparrow}=B_{i\downarrow}=B$ for the Hubbard model on the random regular graph.}\label{f1}
\end{figure}

Let us consider the Hubbard model on a homogeneous quantum interaction graph where $U_i=U, \nu_i=\nu$, and $t_{ij}=t$. In the following, we set $t=1$. For simplicity, we only present the results obtained with real parameters; in fact by the above wave functions and algorithms we find the same behaviors when we take the imaginary parameters into account.  Starting from the MF approximation, in Fig. \ref{f1}, we compare the average charge density in a random regular graph of degree $C=3$ with that in one dimension; the MF approximation qualitatively reproduces the expected phase transitions as the chemical potential $\nu$ increases for fixed $U$. For $U<U_c$, we observe, in turn, an empty phase, a partially filled metallic phase, and a completely filled insulator phase. The figure only displays densities smaller than half-filling; the other part is obtained by the particle-hole symmetry.  For $U>U_c$, a gap opens up at the half-filling density, and we observe another phase transition from the metallic phase to an insulator phase with zero kinetic energy. Notice that, due to the exponential decay of the sign term in this approximation, the model on the random regular graph behaves nearly as the one-dimensional system. The MF approximation correctly predicts the empty-metal transition point where correlations are negligible and underestimates the metal-insulator transition point where strong correlations are responsible for the transition. For the parameters, we find $K_i\simeq K$ and $B_{i\uparrow}\simeq B_{i\downarrow} \simeq B$ in the metallic phase and $K_i\simeq K$  and $B_{i\uparrow}\simeq - B_{i\downarrow} \simeq \pm B$ with a sign that changes from one site to another in the insulating phase. Assuming $K_i=K, B_{i\uparrow}= B_{\uparrow}$, and $B_{i\downarrow}=B_{\downarrow}$, we can easily find the global minimum of the average energy for discrete parameters in a finite region of the parameter space. Up to the half-filling density, we find a paramagnetic solution with $B_{i\uparrow}= B_{i\downarrow}=B$ and after that, a ferromagnetic solution with $B_{i\uparrow}= -B_{i\downarrow}= B$.  
Figure \ref{f1} shows how the optimal parameters change by the chemical potential for the paramagnetic solution.

\begin{figure}
\includegraphics[width=10cm]{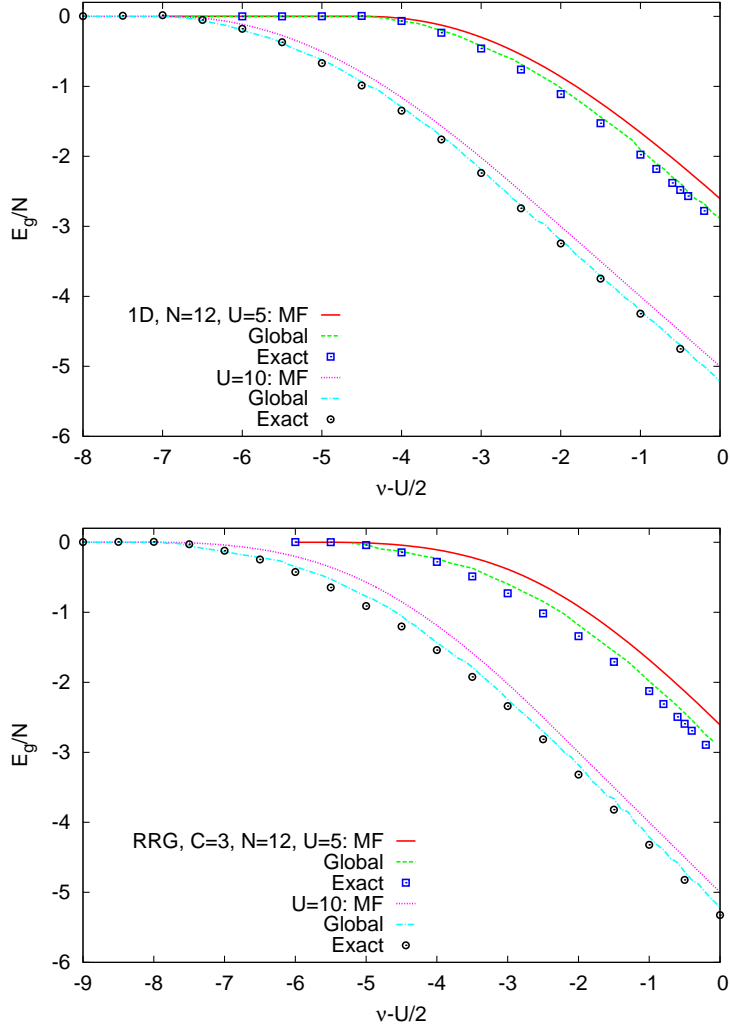}
\caption{The minimum Hamiltonian expectation computed in the MF approximation and the global ansatz compared with the exact ground-state energy computed by the power method, in one dimension (1D), and in a random regular graph (RRG) of degree $C=3$. The data are obtained by the local minimization algorithm and population dynamics ($N_p=100$) followed by the gradient descent algorithm and repeated for $10$ different realizations of the update process.}\label{f2}
\end{figure}

\begin{figure}
\includegraphics[width=10cm]{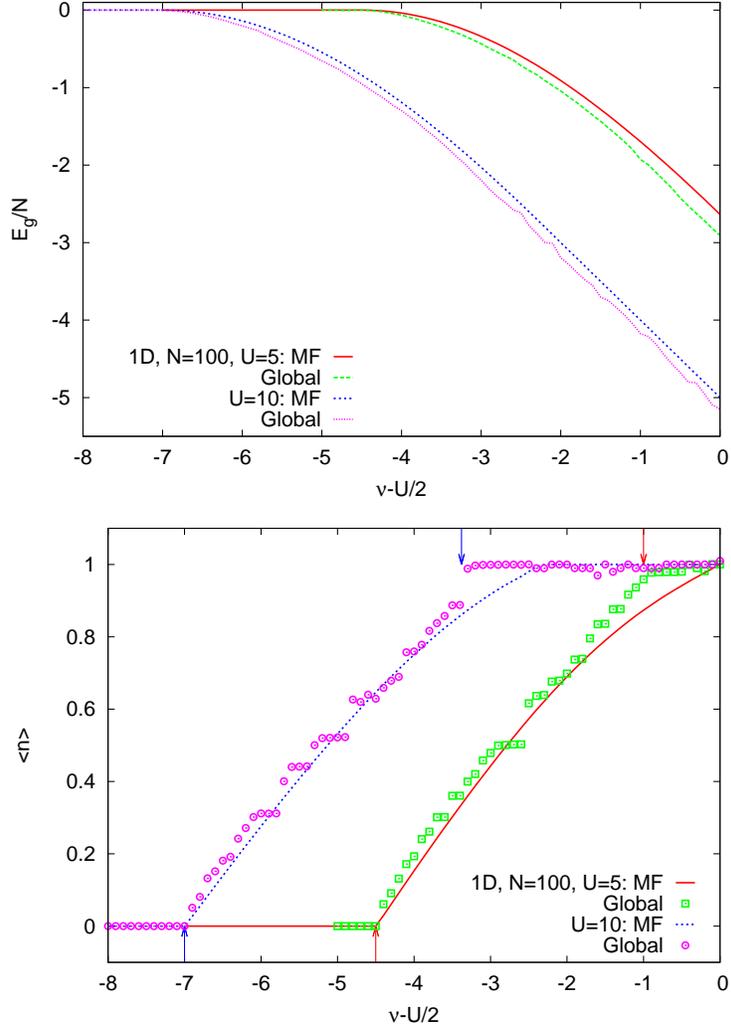}
\caption{The minimum energy and average density computed in the MF approximation and the global ansatz in one dimension and size $N=100$. The data are obtained by the local minimization algorithm and population dynamics ($N_p=100$) followed by the gradient descent algorithm and repeated for $10$ different realizations of the update process.}\label{f3}
\end{figure}

Going beyond the MF approximation, we always obtain smaller energies by adding global interactions other than the local two-body interactions. 
Moreover, even with the global one-body interactions, we obtain results that are comparable with those obtained by the global two-body interactions where it is more difficult to minimize the average energy, and for small and loopy graphs, the approximation errors cancel out the gain from the global two-body interactions. Here, we present the results for the case $\Gamma_{ij\sigma}=0$ where the classical interaction graph has no loops, thus, the Bethe estimation of the Hamiltonian expectation is exact and an upper bound for the ground-state energy; this is the case even if we had the global on-site interactions $\Upsilon_i \xi_{i\uparrow}\xi_{i\downarrow}$ in the classical interaction graph. The global interactions in these wave functions can be considered as effective ones representing higher-order global interactions. The associated wave functions are expected to describe the physical state of the system for $U\gg t$ well. In Fig. \ref{f2} we compare the minimum energies with the exact solutions on small systems computed by the power method using an infinitesimal imaginary time evolution $1-H\tau$ to minimize the Hamiltonian expectation \cite{TC-prb-1990}. Indeed, the power method can also be implemented within the variational formalism where, in each step, one has to project the change in the wave function onto the space of the variational parameters; see Ref. \cite{S-prb-2011} for more advanced methods.

\begin{figure}
\includegraphics[width=10cm]{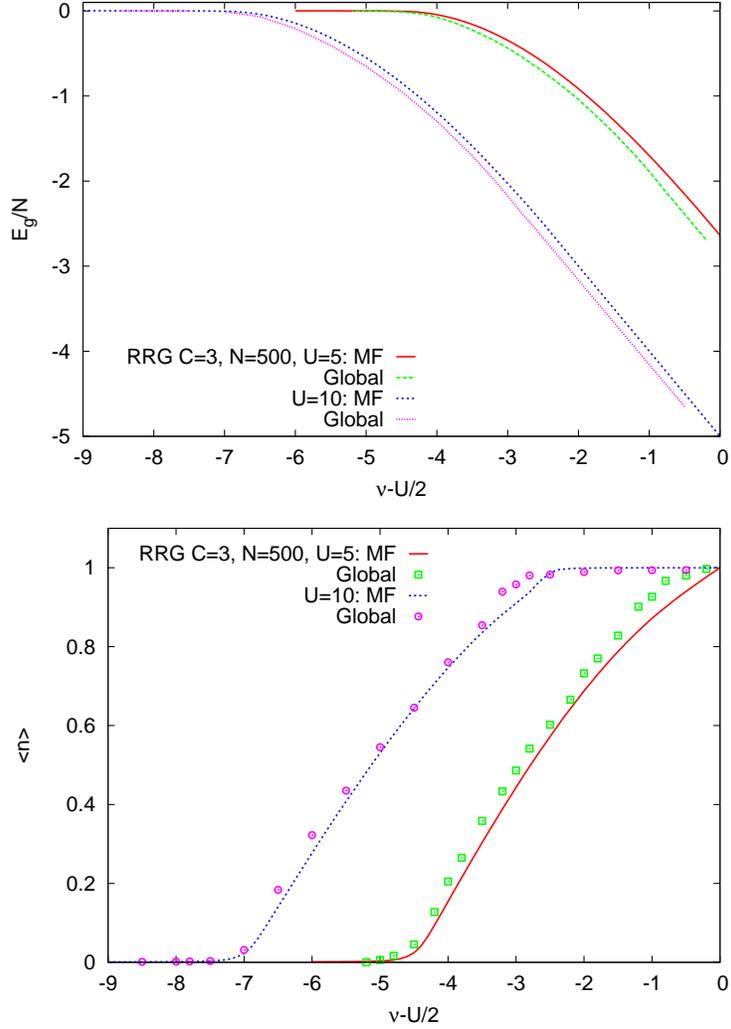}
\caption{The minimum energy and average density computed in the MF approximation and the global ansatz in a random regular graph of degree $C=3$ and size $N=500$. The data are obtained by the local minimization algorithm and population dynamics ($N_p=100$) followed by the gradient descent algorithm and repeated for $10$ different realizations of the update process.}\label{f4}
\end{figure}

\begin{figure}
\includegraphics[width=8cm,height=16cm]{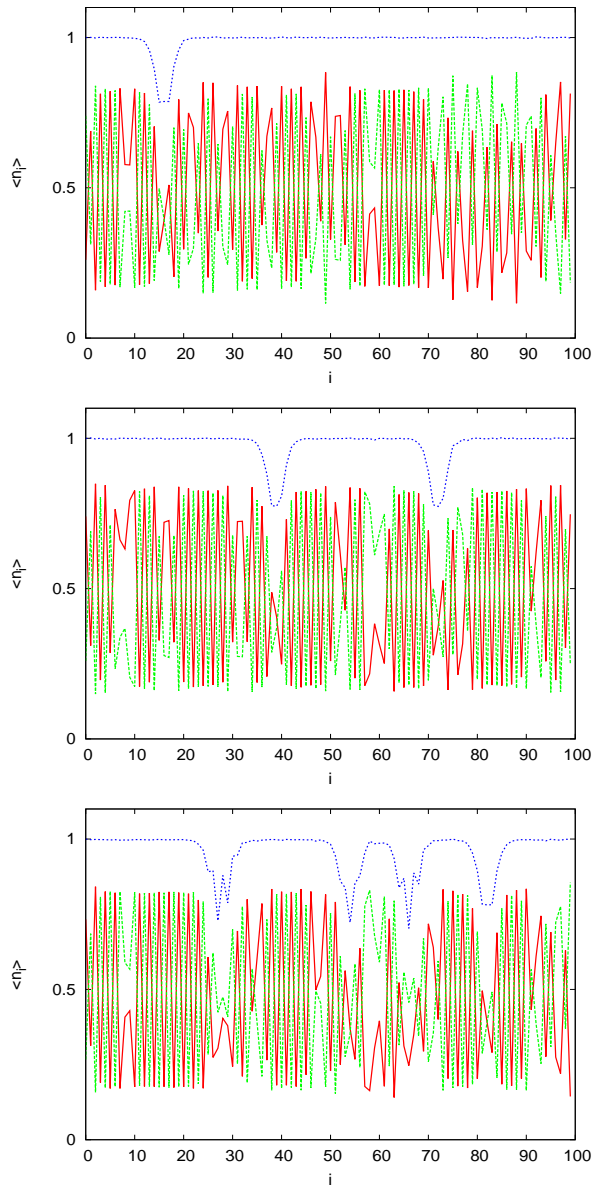}
\caption{The spin and charge density profiles in the global ansatz for different chemical potentials $\nu-U/2=-0.2,-0.5,-1$ (from top to bottom) and $U=5$ in a random regular graph of degree $C=3$ and size $N=100$. The solution is obtained by the population dynamics ($N_p=100$) followed by the gradient descent algorithm and repeated for $10$ different realizations of the update process.}\label{f5}
\end{figure}

Figures \ref{f3} and \ref{f4} display the average charge density for some larger system sizes. In one dimension we are very close to the expected phase transition points in the thermodynamic limit. And in random regular graphs, we observe a shift to smaller chemical potentials for the empty-metal transition point and a shift to larger chemical potentials for the metal-insulator transition point, with respect to the one-dimensional model. One can attribute these shifts to the larger connectivity (here, $C=3$) of the random graphs that provide more degrees of freedom to the particles.

To obtain a picture of the wave functions, in Fig. \ref{f5}, we show the spin and charge density profiles along the ordering backbone of our representation.  In the insulator phase, the spin densities are not frozen on $0$ or $1$ anymore as happens in the MF approximation. In addition, we observe spin density oscillations that were absent in the MF approximation and nonzero kinetic energies in the insulating phase. Before the metal-insulator transition, we observe some charge density holes separating different antiferromagnetic regions. The global one-body parameters $\Theta_{i\sigma}$ alternate between positive and negative signs and are different for the spins up and down.

\begin{figure}
\includegraphics[width=10cm]{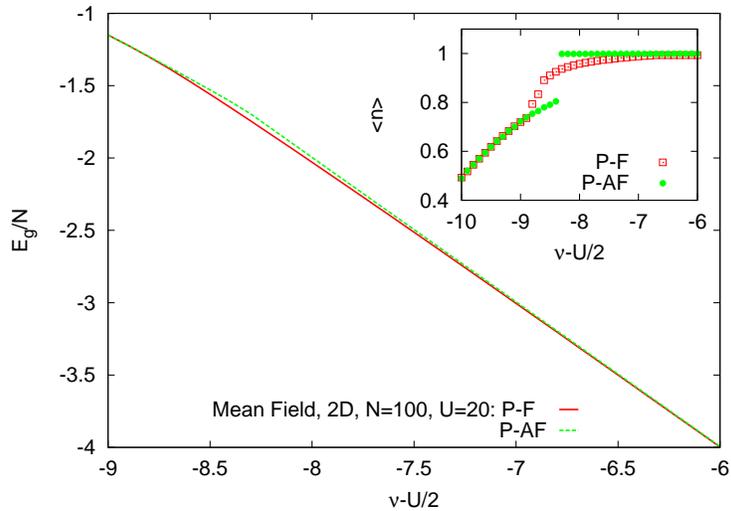}
\caption{The average energy and charge density obtained in the MF approximation on the $2D$ square lattice. The data for paramagnet to ferromagnet (P-F) and antiferromagnet  (P-AF) transitions are obtained by searching for the optimal parameters $(K_i=K, B_{i\uparrow}=B_{\uparrow}, B_{i\downarrow}=B_{\downarrow})$ and $(K_i=K, B_{i\uparrow}=B_{j\downarrow}=B, B_{i\downarrow}=B_{j\uparrow}=\tilde{B})$ (for $i$ in odd and $j$ in even sub-lattices), respectively.}\label{f6}
\end{figure}

\begin{figure}
\includegraphics[width=10cm,height=16cm]{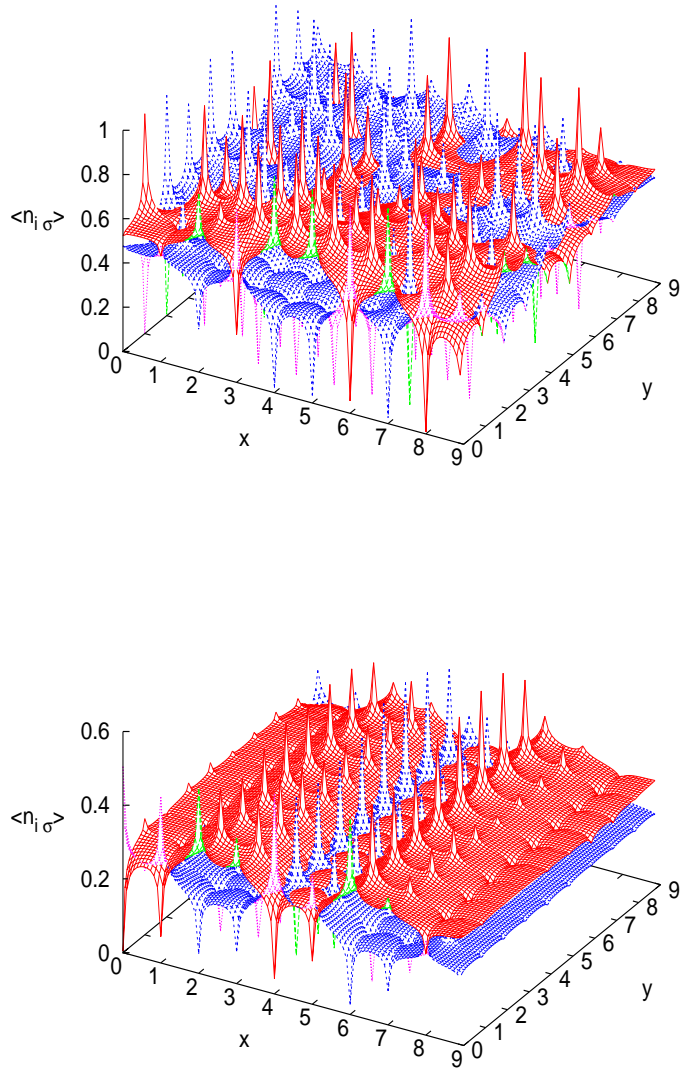}
\caption{The spin density profiles in the global ansatz for chemical potentials $\nu-U/2=-5,-10$ (from top to bottom) and $U=20$ in $2D$ square lattice of size $N=10\times 10$. The solution is obtained by the population dynamics ($N_p=100$) followed by the gradient descent algorithm and repeated for $10$ different realizations of the update process.}\label{f7}
\end{figure}

\begin{figure}
\includegraphics[width=10cm,height=16cm]{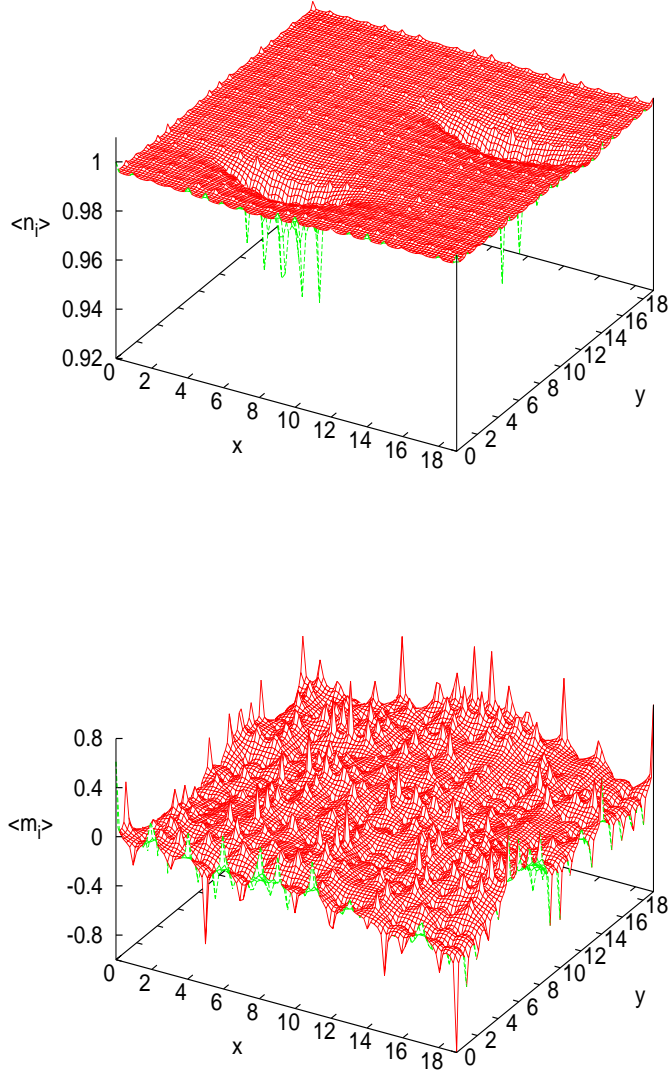}
\caption{The charge density and magnetization ($\langle m_i \rangle=\langle n_{i\uparrow}-n_{i\downarrow} \rangle$) profiles in the global ansatz for $\nu-U/2=-7$ and $U=20$ in $2D$ square lattice of size $N=20\times 20$. The solution is obtained by the population dynamics ($N_p=100$) followed by the gradient descent algorithm and repeated for $10$ different realizations of the update process.}\label{f8}
\end{figure}

Finally, we report some preliminary results for the Hubbard model on the $2D$ square lattice. Figure \ref{f6} shows the average energy and density obtained by the MF approximation when we allow for a phase transition from the paramagnetic (P) phase to the ferromagnetic (F) and antiferromagnetic (AF) phases. The former transition happens before the latter one, and the difference between the two energies decreases as one approaches the half-filling density where the AF phase has a smaller energy. Adding the global one-body interactions, we find a transition from the P phase to a mixed phase of F and AF regions, see Fig. \ref{f7}; the system is more ferromagnetic (antiferromagnetic) for smaller (larger) chemical potentials. Figure \ref{f8} displays the charge and spin profiles close to the half-filling density. We observe tendencies towards the holes condensation \cite{V-prb-1974} where a ferromagnetic phase of small density is separated from an antiferromagnetic phase of higher density. 

\begin{figure}
\includegraphics[width=10cm]{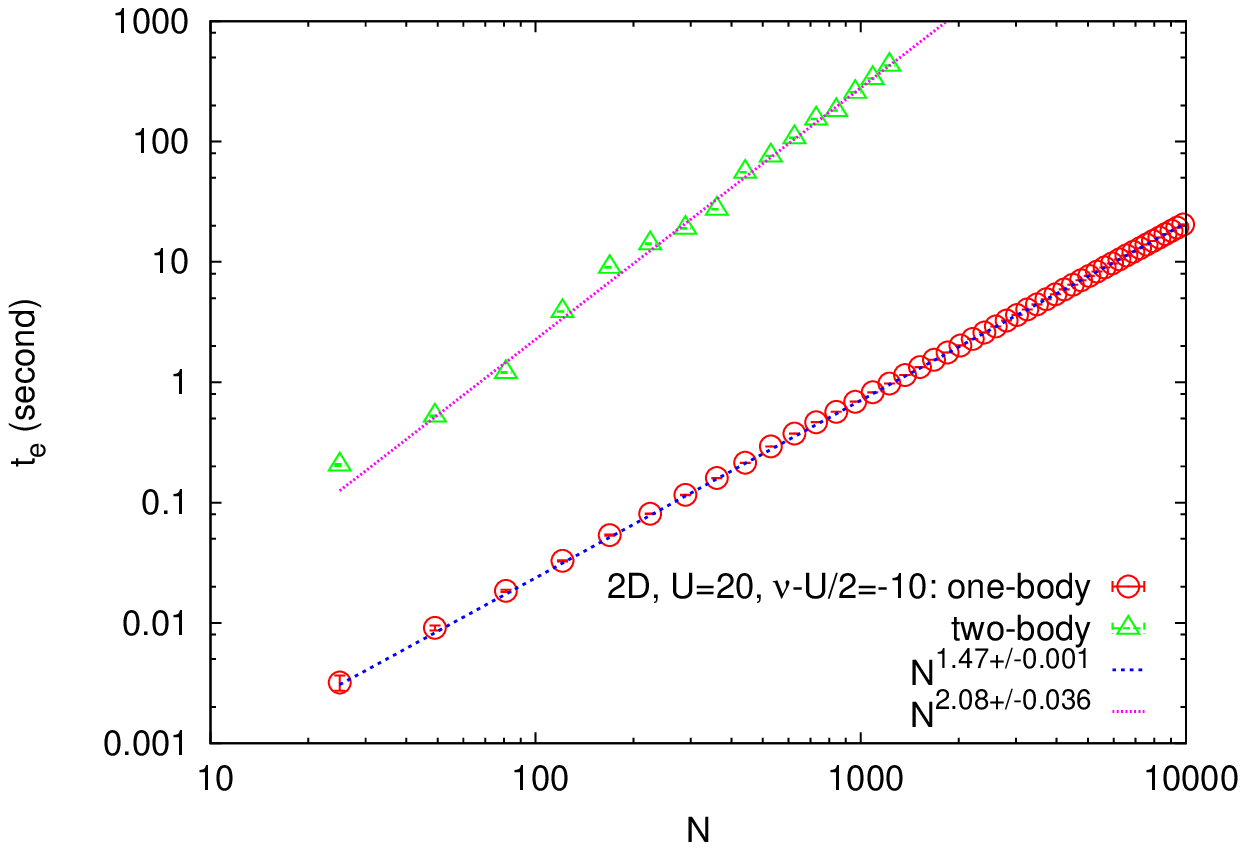}
\caption{The typical CPU time $t_e$ we need to compute the Hamiltonian expectation and the average local observables in the $2D$ square lattice with the global one-body  interactions given the set of parameters $K_i,B_{i\sigma},\Theta_{i\sigma} \in (-5,5)$. We also show the computation time for a diluted lattice, where on average $10$ percent of the global two-body interactions are present in addition to the ones in the ordering backbone, for $\Gamma_{ij\sigma}\in (-1,1)$. For the one-body interactions the BP algorithm always converges but with the two-body interactions the algorithm may not converge. In the latter case we limited the maximum number of BP iterations to $T_{max}=500$. The average time has been obtained for $100$ realizations of the parameters.}\label{f9}
\end{figure}

In Fig. \ref{f9}, we display the time $t_e$ we need to compute the Hamiltonian expectation and the local average quantities in the $2D$ square lattice with the global one-body interactions, given the set of parameters $K_i,B_{i\sigma}$, and $\Theta_{i\sigma}$. The computation time for updating the population of the parameters (one sweep) is $N_pt_e$, and, in practice, we need a few hundred sweeps of the updates to reach the stationary state.
We are reminded that, as long as one only considers the global one-body interactions, the algorithm gives the exact Hamiltonian expectation and, therefore, a good upper bound for the ground-state energy. For comparison, we also show the computation time for a diluted lattice where a small fraction of the global two-body interactions are present in addition to the ones in the ordering backbone.

\section{Discussion}\label{S4}
The Hamiltonian expectation can be computed by an efficient and distributive message-passing algorithm, which is asymptotically exact on random and sparse interaction graphs. To obtain a good estimation of the average fermion sign, we have to work with the global interactions in the classical system. Unfortunately, 
this makes the average energy a nonlocal function of the variational parameters, resulting in an optimization problem that is not amenable anymore to local message-passing algorithms. Moreover, the performance of any optimization algorithm strongly depends on the quality of the estimated average energy or the approximation method that is used to take the average of the energy function. The study can systematically be improved in both directions by considering more accurate inference algorithms using generalized Bethe approximations \cite{K-pr-1951,YFW-nips-2001} or incorporating replica-symmetry-breaking and more sophisticated optimization algorithms to find the optimal variational parameters.

\acknowledgments
We are grateful to A. Montorsi, M. M\"{u}ller and S. Sorella for helpful discussions. Support from ERC  Grant No. OPTINF  267915 is acknowledged.

\end{document}